\documentclass{aa}
\usepackage{graphics}
\begin{document}

\thesaurus{(08.01.2; 08.03.5; 08.06.1; 08.09.2; 13.25.5)}

\title{A {\it Beppo}SAX Observation of HD\,9770: a Visual Triple System
Containing a Recently Discovered Short-Period Eclipsing Binary}

\author{G. Tagliaferri\inst{1} 
	\and S. Covino\inst{1} 
 	\and G. Cutispoto\inst{2}
	\and R. Pallavicini\inst{3}}
\offprints{G. Tagliaferri}
\institute{Brera Astronomical Observatory, Via Bianchi 46, I-22055 Merate, 
	Italy 	
	\and Catania Astrophysical Observatory, V.le A.\,Doria 6, I-95125 
	Catania, Italy
	\and Palermo Astronomical Observatory, P.za del Parlamento 1, 
	I-90134 Palermo, Italy}

\date{Received date / accepted date}

\titlerunning{A SAX Observation of HD\,9770}
\authorrunning{Tagliaferri et al.\ }

\maketitle

\begin{abstract}
We have studied the coronal X-ray emission of the recently discovered 
short-period eclipsing binary HD\,9770 with the {\it Beppo}SAX satellite.
The data from the Low Energy and Medium Energy Concentrator Spectrometers
(LECS \& MECS) onboard {\it Beppo}SAX allow studying the spectrum
of this star from 0.1 to 8 keV, confirming that this is a very active
coronal source, with strong flaring activity. The X-ray emission most
likely originates from the eclipsing binary itself, rather than from 
the other visual component of the system. The X-ray light curves
could be modulated with the orbital period of the eclipsing binary,
with a hint for a different orbital modulation of the cooler
and hotter plasma. The X-ray spectrum is characterized by
hot plasma, with the Fe K complex at 6.7 keV clearly detected in the
MECS spectrum, and it is well fitted by a 2-temperature 
optically thin plasma model with low metal abundances ($\sim 0.3\,Z_\odot$).
These results are in line with those found for many other active stars.
As expected, during the flare the X-ray emission is dominated by
hotter plasma with a temperature $> 4$ keV. There is an indication 
that the metal abundance may be somewhat higher during the flare.

\end{abstract}

\section{Introduction}
 
\begin{figure}
\begin{center}
{\resizebox{\hsize}{!}{\includegraphics{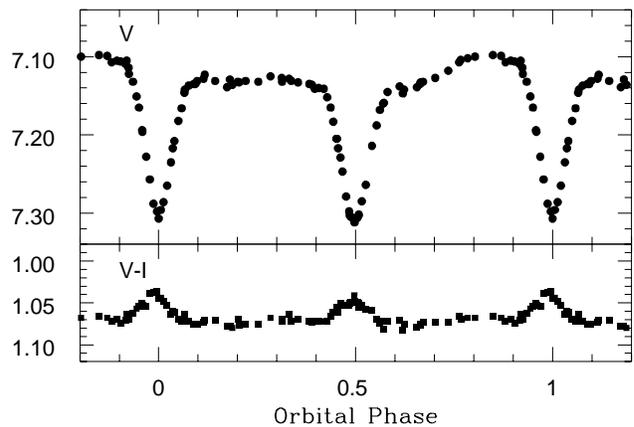}}}
\end{center}
\caption{HD\,9770 V-band light curve and $V - I$ color obtained over the
interval 7--25 Dec 1996, folded with the period of P=0.4765318\,days.}
\label{fig:opt_lc}
\end{figure}

\object{HD\,9770} 
(\object{SAO\,193189}; \object{CD--30529}; \object{PPM\,244106}; 
\object{GSC\,\-064\-28\,01616}; \object{BB\,Scl}) is a very interesting
nearby (24\,pc as determined by the {\it Hipparcos} satellite) 
visual triple system with a total magnitude V\,=\,7.1. Components AB and C 
were classified as K3V 
and M2V, respectively (Edwars 1976; Gliese 1969). 
Components A and B are separated by 0.17\,arcsec, while the much weaker 
component C is at 1.44\,arcsec from A + B (Gliese 1969).
Cutispoto et al. (\cite{CKM97}), on the basis of multicolour UBV(RI)$_c$
photometry, have classified
component A as a K1/2V type star; they also showed that the B component is
itself a short period ($\sim 0.48$\,days) eclipsing binary formed by
two nearly identical stars of spectral type K4/5\,V and K5\,V, respectively. 
It shows two almost identical primary and secondary eclipses that last
for about 70 minutes each, and out-of-eclipse variability that indicates
that \object{HD\,9770} belongs to the BY Dra type of variable stars 
(Fig.\,\ref{fig:opt_lc}).
The conclusion that component B is indeed the eclipsing binary comes from 
the peculiar behavior of the colour indices at both primary and secondary
eclipses. In fact at both minima, the star is bluer indicating that the
cooler eclipsing component (B$_1$ + B$_2$) is contributing less to the total
integrated light of the A + B system (component C is too faint to generate
the observed light variations). More recently Cutispoto et 
al. (1999) revised the spectral classification of the A, B$_1$ and B$_2$
components to K0/1\,V, K3\,V and K3/4\,V, respectively, in more agreement
also with the {\it Hipparcos} distance.

\object{HD\,9770} was clearly detected during the all-sky EUV surveys
obtained with both the WFC on board the ROSAT satellite (Pounds et al.\  
\cite{PAB93}, Pye et al.\  \cite{PMA95}; \object{RE\,0135--295}) and with the 
EUVE satellite (Malina et al.\  \cite{MMA94}; \object{EUVE\,J0135}). It was 
also detected in the ROSAT all-sky 
survey (RASS, Snowden \& Schmitt \cite{SS90}; \object{1RXS\,J013500.7--295430}) 
with an average PSPC count rate of $\sim 3.5$ cts s$^{-1}$
in the spectral band 0.1--2.4 keV. The star was observed by the PSPC
in individual scans about 30\,s long spaced by 96 min, covering more 
than 2\,days, with an additional single scan obtained after a gap of about
160\,days. The source was clearly variable, with a mean X-ray 
luminosity of $L_{\rm x}=1.6 \times 10^{30}$
${\rm erg}~{\rm sec}^{-1}$ (Cutispoto et al.\  \cite{CKM97}).
During the two-day observation, the X-ray luminosity ranged
from $\sim 9 \times 10^{29}$ to $2.3 \times 10^{30}$ erg s$^{-1}$, with
a single flaring data point as high as $3.7 \times 10^{30}$ erg s$^{-1}$.
In all these X-ray and EUV observations, the triple visual system was
observed as a single unresolved source. The high X-ray luminosity of the
source strongly suggests that most of the X-ray flux originated from the
eclipsing binary, due to its short period and more efficient dynamo activity.
The contribution of components A and C is expected to be at least one order
of magnitude less if these components behave as normal K and M field dwarfs
(Schmitt et al. 1995). K and M stars could provide a meaningful contribution
to the X-ray flux only if they were very young. This does not seem to be
the case, since in optical spectra obtained recently by us at ESO we do not 
detect any lithium line (with an upper limit of $\mbox{EW} < 12$\,m\AA\ for the
6707.8\,\AA\ lithium line) and there are no indications for a very young age of 
this system (actually, we find [Fe/H]$\sim -0.5$, see Sect. 4). 

Thanks to its short period, we monitored $\sim 3.5$ contiguous orbital 
cycles of \object{HD\,9770} with the {\it Beppo}SAX satellite. 
The purpose was to
study the short-term variability of this source in search of 
rotational/orbital modulation as well as of eclipses, while at the same
time determining the spectral properties of the source.
Although {\it Beppo}SAX has a lower energy resolution, typical of 
gas-scintillation proportional counters, than the ASCA CCD detectors,
its extended spectral range, from 0.1 to 10 keV for coronal sources, is well 
suited to study the coronal temperature and emission measure distribution
(see also Favata et al. \cite{FMPS98}).  
Here we present the spectral results together with the optical monitoring
of the {\it Beppo}SAX observation of \object{HD\,9770}.

\section{Observations}

The {\it Beppo}SAX satellite carries on board various X-ray detectors
covering a very large energy band from 0.1 to 300 keV (Boella et al.\  
\cite{BBP97a}). For the study of typical coronal sources the most 
suitable detectors are the Low Energy Concentrator Spectrometer (LECS, Parmar 
et al.\  \cite{PMB97}) and the three Medium Energy Concentrator Spectrometers 
(MECS, Boella et al.\  \cite{BCC97b}). The LECS has a wide energy range, 
0.1--10 keV, and good spectral resolution, comparable to CCD detectors at low 
energies where it fills the gap between EUVE and ASCA. On the contrary, the 
three MECS detectors cover only the 1.7--10 keV energy range, but they
have an effective area about three times larger than the LECS, thus
allowing the study of the Fe K complex at $\sim 6.7$ keV much more 
effectively than with the LECS, although at a lower spectral resolution
than with ASCA.

{\it Beppo}SAX observed \object{HD\,9770} on December 7--9, 1996 for 
about 42 hours, resulting in 40 ks and 83 ks of observing time in the LECS
and MECS detectors, respectively. The difference is due to the LECS being
operated only when the spacecraft was in the Earth shadow. The data
analysis was based on the linearized, cleaned event files obtained from
the {\it Beppo}SAX Science Data Center (SDC) on-line archive (Giommi \& Fiore 
\cite{GF97}). Light curves and spectra were accumulated with 
the FTOOLS package (v. 4.0), using an extraction 
region of 8.5 and 4\,arcmin radius for the LECS and MECS, respectively.
At low energies the LECS has a broader Point Spread Function (PSF) than the 
MECS, while above 2\,keV the PSFs are similar. The adopted regions provide 
more than 90\% of the source counts at all energies both for the LECS and MECS.
The LECS and MECS background is 
low, but not uniformly distributed across the detectors, on the other
hand it is rather stable. For this reason, it is better to evaluate the
background from blank fields, rather than in annuli around the source region.
Thus, after having checked that the background was stable during the whole
observation by analyzing a light curve extracted from a source-free region, 
for the spectral analysis we used the background files accumulated
from long blank field exposures and available from the SDC public ftp site
(see Fiore et al. \cite{FGG99}, Parmar et al. \cite{Petal98}).
We did not subtract the background from the light curves given that
it is stable and it amounts to only $\sim 1/6$ and $\sim 1/5$ of the MECS and 
LECS source flux, respectively.

The spectral analysis was performed with the XSPEC 10.0 package,
using the response matrices released by the SDC in September 1997. We binned
the spectra using the rebinning template files provided by the SDC. These files
contain a specific rebinning to sample the instrument resolution with the
same number of channels, three in our case, at all energies (i.e. the rebinning
factor is not constant with energy). For the spectral analysis, the LECS data 
have been considered only in the range 0.1--4 keV, due to still unsolved 
calibration problems at higher energies (Fiore et al. \cite{FGG99}). To fit 
the LECS and MECS spectra together, one has to introduce a constant rescaling 
factor to account for uncertainties in the inter-calibration of the instruments. 
The acceptable values for this constant is within 0.7 and 1 (Fiore et al. 
\cite{FGG99}). The best fit value in our case using the full datasets is 0.83, 
in full agreement with the 
acceptable range. Thus we kept this constant value fixed to 0.83 in all our 
spectral analysis \footnote{A preliminary analysis of this {\it Beppo}SAX 
observation of \object{HD\,9770} (Tagliaferri et al.
\cite{TCCP98a,TCCP98b}) gave slightly different results with respect to 
those reported here. This is due to the adoption for the present analyses of 
new calibration files provided by the SDC.}. For a full description of the
{\it Beppo}SAX data analysis see Fiore et al. (\cite{FGG99}).

Additional spectral information can be obtained by the Phoswich Detector 
System (PDS, Frontera et al.\ \cite{FCF97}) on board {\it Beppo}SAX 
albeit this detector was designed to provide the maximum sensitivity at 
energies somewhat higher than for typical coronal sources. The energy range 
covered by the PDS is 15 to 300 keV and the experiment can perform sensitive 
spectral and temporal studies over this energy range. Hard ($> 10$ keV) X-ray 
emission has been detected by the PDS from large flares on active stars 
(e.g. Favata 1998, Pallavicini and Tagliaferri 1998, Pallavicini et al. 
\cite{PTM99}); 
however, no X-ray emission was detected by the PDS for \object{HD\,9770}, not 
even during flares, so we will not consider this instrument anymore in the 
following discussion.

The photometric optical data presented in Fig.\,\ref{fig:opt_lc}, are part 
of the {\it UBV(RI)$_c$} observations carried out
in the period 7--25 Dec, 1996 at the European Southern Observatory
(La Silla, Chile). We used the 50\,cm ESO telescope equipped with a
single-channel photon-counting photometer, a thermoelectrically cooled R943--02
Hamamatzu photomultiplier and standard ESO filters matching the $UBV(RI)_c$
system. Details on the observation and reduction procedures can be found in
Cutispoto (1995). 
These data allowed us to extend the time window over which to perform
the period search. In this way we were able to obtain a new, more
accurate, period of $0.4765318 \pm 0.0000012$\,days (to be compared with the
previous period of $0.476533 \pm 0.000033$\,days, obtained by Cutispoto et
al. 1997).

\begin{figure}
\begin{center}
\begin{tabular}{c}
{\resizebox{\hsize}{!}{\includegraphics{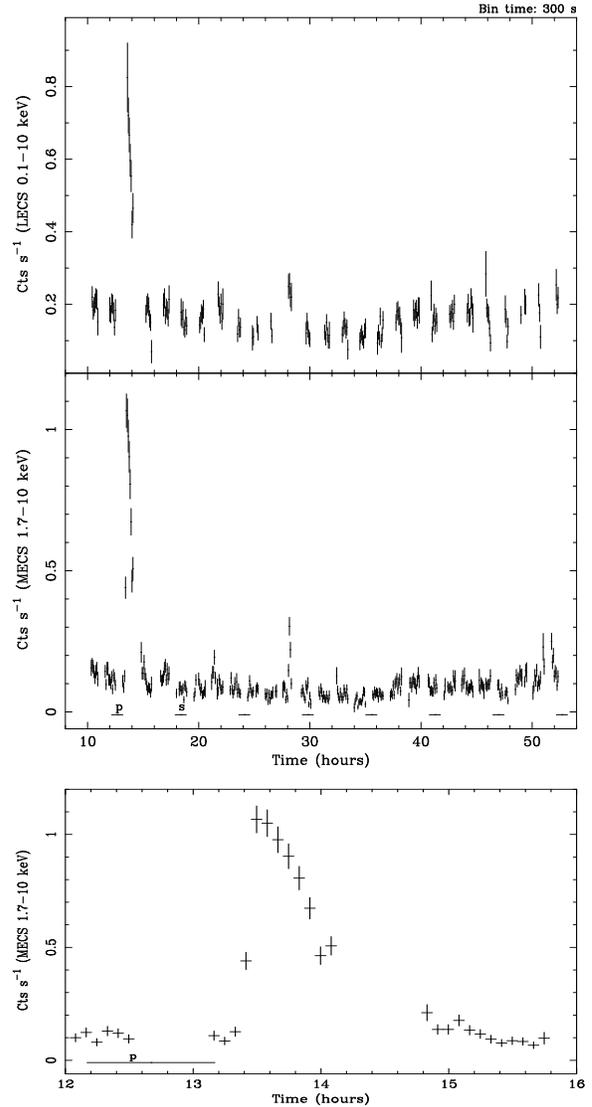}}} \\
\end{tabular}
\end{center}
\caption{SAX LECS and MECS light curves for \object{HD\,9770} (top panel) and 
the region around the main flare for the MECS (bottom panel), binned over time 
bins of 300\,s. The observation started on December 7, 1996. The dashes below 
the MECS light curves show the position and duration of the primary and 
secondary optical eclipses.}
\label{fig:lc}
\end{figure}

\section{Results} 

\begin{figure}
\begin{center}
\begin{tabular}{c}
{\rotatebox{270}{\resizebox{17cm}{!}{\includegraphics{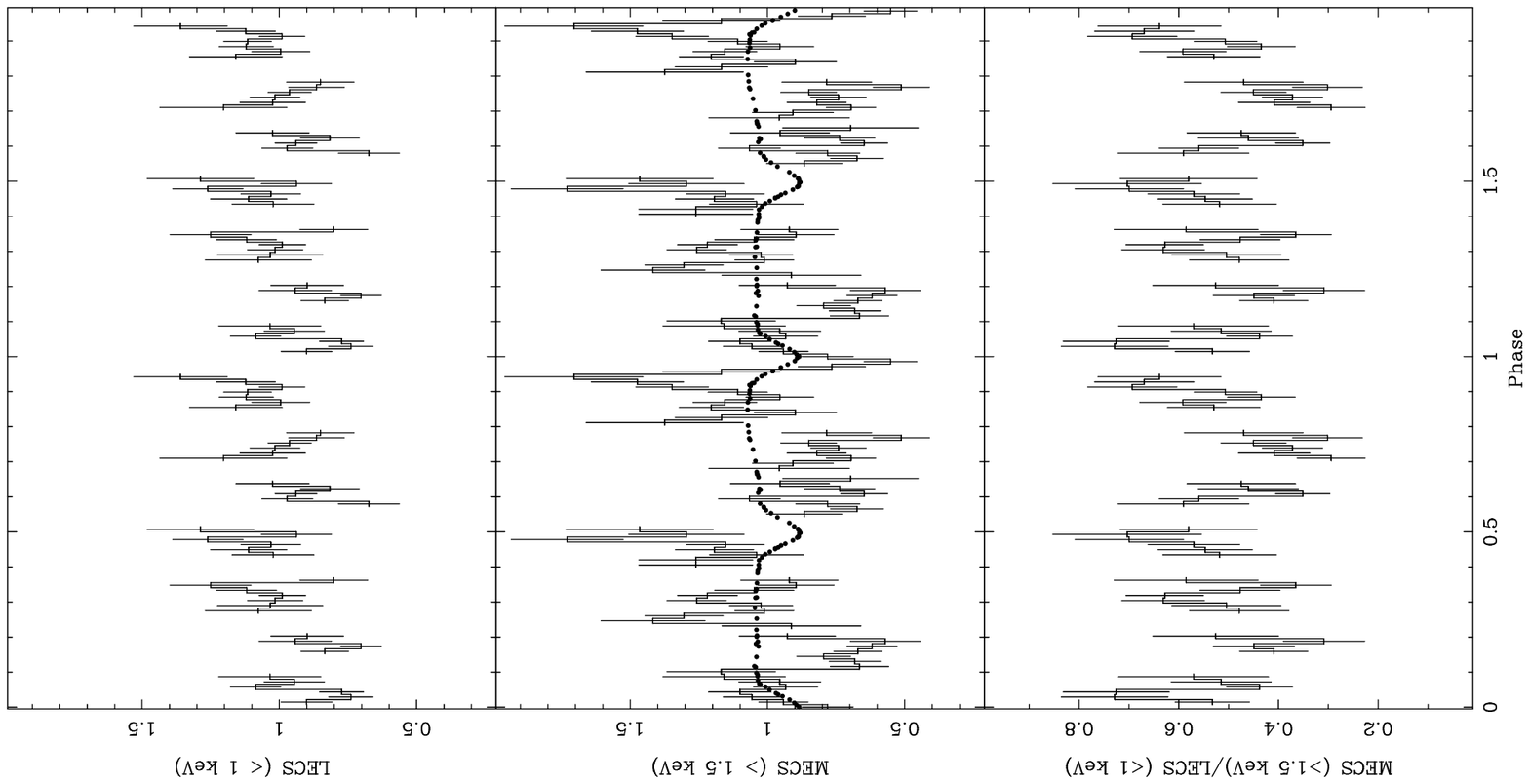}}}}\\
\end{tabular}
\end{center}
\caption{SAX MECS (top panel), LECS (middle panel), and MECS/LECS light 
curves for HD\,9770, binned over time bins of 600\,s. The folded, with the
optical period, light curves have been obtained within the following time 
windows: December 7, 1996; UT 10:17-13:12, 14:38-27:21, 28:48-50:24. The 
curves, but the hardness ratio, are normalized dividing by the average source 
intensity. The relative energy bands used for the MECS and LECS data are also 
given.}
\label{fig:foldedlc}
\end{figure}

In Fig.~\ref{fig:lc} we plot the total MECS light curve of \object{HD\,9770}.
Times are measured starting from December 7, 1996 \mbox{00:00} UT.
A strong flare is clearly detected at the beginning of the observation at
phase $\sim 0.2$. A second smaller flare was detected about 15 hours later,
during the second observed cycle of the binary, at phase $\sim 0.5$. 
Another flare was probably just missed, due to Earth occultation, at the 
end of the observation, during the fourth cycle, again at phase 0.5 (note
that the primary and secondary eclipses are centered at phases 0 and 0.5
and that the whole observation in Fig.\,\ref{fig:lc}  
comprises $\sim 3.5$ orbital cycles). The light curves were then folded
(Fig.\,\ref{fig:foldedlc}) using the new ephemerides derived by us
with optical observations (see previous section and Fig.\,\ref{fig:opt_lc}).
A narrow time window around the main flare was removed, otherwise 
the flare would have dominated the folded light curve, while
we are interested instead in the presence of an orbital modulation. 
We also removed the two smaller flares detected at $\sim 28$ and 51 hours 
(see Fig.\,\ref{fig:lc}). For the exact time windows see caption of 
Fig.\,\ref{fig:foldedlc}.
The top and the middle panels of Fig.\,\ref{fig:foldedlc} show the MECS and 
LECS folded light curves, respectively. If present, an orbital 
modulation would imply that the size of the coronal loop structures are smaller 
than or at most comparable to the stellar radius and that they are distributed 
not uniformly across the stellar disk.
Unfortunately, due to the large error bars, a period search did not give 
conclusive answers, even if we assume the known optical period. In fact,
with our statistics, we could only detect an orbital modulation if this would
be greater than $50\%$ ($3\,\sigma$ detection). 
Nevertheless, simply from a visual inspection of the folded light curve with the
optical period, a modulation appears to be present. This seems more pronounced 
in the harder band (MECS, $> 1.5$\,keV), where two maxima can be identified at
orbital phases between $0.4 \div 0.5$ and $0.8 \div 0.9$. To put this on more 
solid statistical ground, we fitted the LECS ($< 1$\,keV), MECS ($> 1.5$\,keV) and 
hardness ratio (MECS [$> 1.5$\,keV]/LECS [$< 1$\,keV]) folded light curves
with a constant in order to check the amount of variability. 
The reduced $\chi^2$ are 1.45, 2.79, 1.36 for the LECS, MECS and hardness 
ratio folded light curve, respectively. In all three cases the 
hypothesis of a constant curve can be rejected with a probability greater than 
99.9\%, however the variability is clearly more pronounced in the MECS.
For this analysis
we selected only simultaneous LECS and MECS data, to avoid bias introduced 
by the larger gaps in the LECS light curve. We also excluded the flare activity 
using the same windows reported in the caption of Fig.\,\ref{fig:foldedlc}. 
However, we plot in Fig.\,\ref{fig:foldedlc} the MECS folded light curve
obtained using the full MECS data set, but the flare windows, in order to show 
to the reader as much information as possible. We also fitted 
the curves with a constant plus a sinusoidal model to verify if it can 
reproduce the observed variability. The reduced $\chi^2$ are
1.18, 1.51, 1.03 for the three curves, respectively. A formal F test shows
that the improving of the $\chi^2$ is significant at a level greater than
99\%. Clearly, there is more variability in the folded light curves than what
is reflected by this simple model, but the improvement is significant. 
In the MECS folded light curve, which is the one with more information,
the primary eclipse at phase 0 seems detected, immediately after
the second maximum. 
The secondary eclipse is not detected, actually at this position we have no
data due to the windowing. Otherwise, we would have a maximum in the flux
at phase $\sim 0.5$ due to the small flare seen during the second cycle and to the one 
partly missed at the end of the observation, during the fourth cycle.
This probably implies that there is an active region on the primary star
rotating in front while it is eclipsing the secondary. The bottom panel
of Fig.\,\ref{fig:foldedlc} finally shows the hardness ratio MECS/LECS;
a softening of the X-ray spectrum during the minima seems to be present.
This would indicate that the hotter coronal plasma occupies a smaller volume.

For the spectral analysis the whole dataset was subdivided into
two parts. The ``flare'' set, composed only of events related to the main flare
starting at $\sim \mbox{13:25}$ hours of the observing time, 
and the ``out of flare'' dataset with all events but those associated 
with the main flare. Note that the count rate and the duration of the flare
were not sufficiently high/long to allow time-resolved spectroscopy 
throughout the flare. For this reason, we do not attempt a modeling of this
event.

\begin{table*}
\begin{center}
\begin{tabular}{|l|cccccc|}
\hline
& $KT_1$        & EM$_1$ & $KT_2$ & EM$_2$ & $Z$ & $\chi^2_\nu$ \\
& (keV) & $10^{52}$\,cm$^{-3}$ & (keV) & $10^{52}$\,cm$^{-3}$ & $Z/Z_\odot$ & \\
\hline
& & & & & &\\
Flare RS & $0.82\pm^{0.27}_{0.30}$ & $2.95$ &
$4.26\pm^{1.68}_{0.85}$ & $15.55$ & $0.51\pm^{0.25}_{0.26}$ &
0.85 \\
Flare MK & $0.80\pm^{0.42}_{0.30}$ & $3.00$ &
$4.23\pm^{1.60}_{0.83}$ & $18.31$ & $0.52\pm^{0.40}_{0.27}$ &
0.86 \\
& & & & & &\\
Out of Flare RS & $0.84\pm^{0.04}_{0.06}$ & $3.43$ &
$1.98\pm^{0.40}_{0.28}$ & $2.52$ & $0.30\pm^{0.09}_{0.07}$ &
0.88 \\
Out of Flare MK & $0.71\pm^{0.09}_{0.09}$ & $2.43$ &
$1.75\pm^{0.24}_{0.15}$ & $3.36$ & $0.35\pm^{0.11}_{0.09}$ &
0.82 \\
& & & & & &\\
\hline
\end{tabular}
\end{center}
\caption{Two-temperature RS and MK best fit parameters. Spectra are analyzed 
considering both the strong flare and the whole observation but this 
flare. The energy bins are 110 for each dataset and the corresponding 
degrees of freedom are 105. Errors are computed at the 90\% confidence
level assuming three interesting parameters ($\Delta \chi^2 = 6.21$).}
\label{tab:spec}
\end{table*}

The spectral fits were performed with the optically thin plasma models of
Raymond \& Smith (\cite{RS77}; hereafter RS) and Mewe et 
al.\  (\cite{MKL96a}; hereafter MK) models, as implemented inside XSPEC. 
One- and two-temperature models were assumed with metal abundances $Z$
either fixed to the solar value or varied in a fixed proportion with respect
to solar. The interstellar absorption $N_{\rm H}$
was also included in the fit. We first left $N_{\rm H}$ free to vary in the
fit procedure, obtaining a best-fit value of the order of $3 \times
10^{19}$\,cm$^{-2}$. This value is quite high for a star whose distance
is 24\,pc (as measured by Hipparcos); for instance if 
we apply the relation ${\rm H} \sim 0.07$\,cm$^{-3}$ from Paresce (\cite{P84})
with the above distance we get $N_{\rm H} \sim 5.2 \times 10^{18}$\,cm$^{-2}$.
High value of $N_{\rm H}$ are obtained also for other {\it Beppo}SAX
observations of coronal sources: VY\,Ari, II\,Peg, UX\,Ari
and AB\,Dor (Favata et al. \cite{FMP97b}, Tagliaferri et al.1999b, Pallavicini 
et al. 1999). This
anomaly could either be due to a problem in the calibration of the
LECS detector below 0.5 keV (which is where the $N_{\rm H}$ is estimated
in the case of values lower than $\sim \times 10^{20}$ cm$^{-2}$) or
to a wrong modeling, whose deficiency is revealed by the large energy
range covered by the {\it Beppo}SAX data.
In order to minimize the number of free parameters and also the influence
that $N_{\rm H}$ could have on the derived metallicities at the low resolution
of our detector (a higher $N_{\rm H}$ would mimic the same effect as a lower 
metallicity, although for the $N_{\rm H}$ range here involved this is practically 
negligible), we fixed the value of $N_{\rm H}$ to $5\times 10^{18}$\,cm$^{-2}$. 
Moreover, for the evaluation of the count errors, we adopted the 
approximation of Gehrels (\cite{G86}) for data following the Poisson 
statistics, instead of the less accurate Gaussian approximation.

\begin{table*}
\begin{center}
\begin{tabular}{|l|ccc|c|}
\hline
& flux$_{0.1-10\,{\rm keV}}$ & flux$_{0.1-2.5\,{\rm keV}}$ &
flux$_{0.5-10\,{\rm keV}}$ & $L_{0.1-2.5\,{\rm keV}}$ \\
& erg\,cm$^{-2}$\,sec$^{-1}$ & erg\,cm$^{-2}$\,sec$^{-1}$ &
erg\,cm$^{-2}$\,sec$^{-1}$ & erg\,sec$^{-1}$ \\
\hline
& & & & \\
Flare & $6.5\times10^{-11}$ & $4.2\times10^{-11}$ & $5.4\times10^{-11}$ & $2.9\times10^{30}$ \\
& & & & \\
Out of Flare &$1.3\times10^{-11}$ &$1.2\times10^{-11}$ &$9.9\times10^{-12}$ & $8.2\times10^{29}$ \\
& & & & \\
\hline
\end{tabular}
\end{center}
\caption{Two-temperature best fit LECS fluxes in three different bands
and X-ray luminosities in the ROSAT 0.1--2.5\,keV band. The reported fluxes 
are the same for both the RS and MK models.}
\label{tab:fluxes}
\end{table*}

\begin{figure}
\begin{center}
\begin{tabular}{c}
{\rotatebox{270}{\resizebox{13cm}{!}{\includegraphics{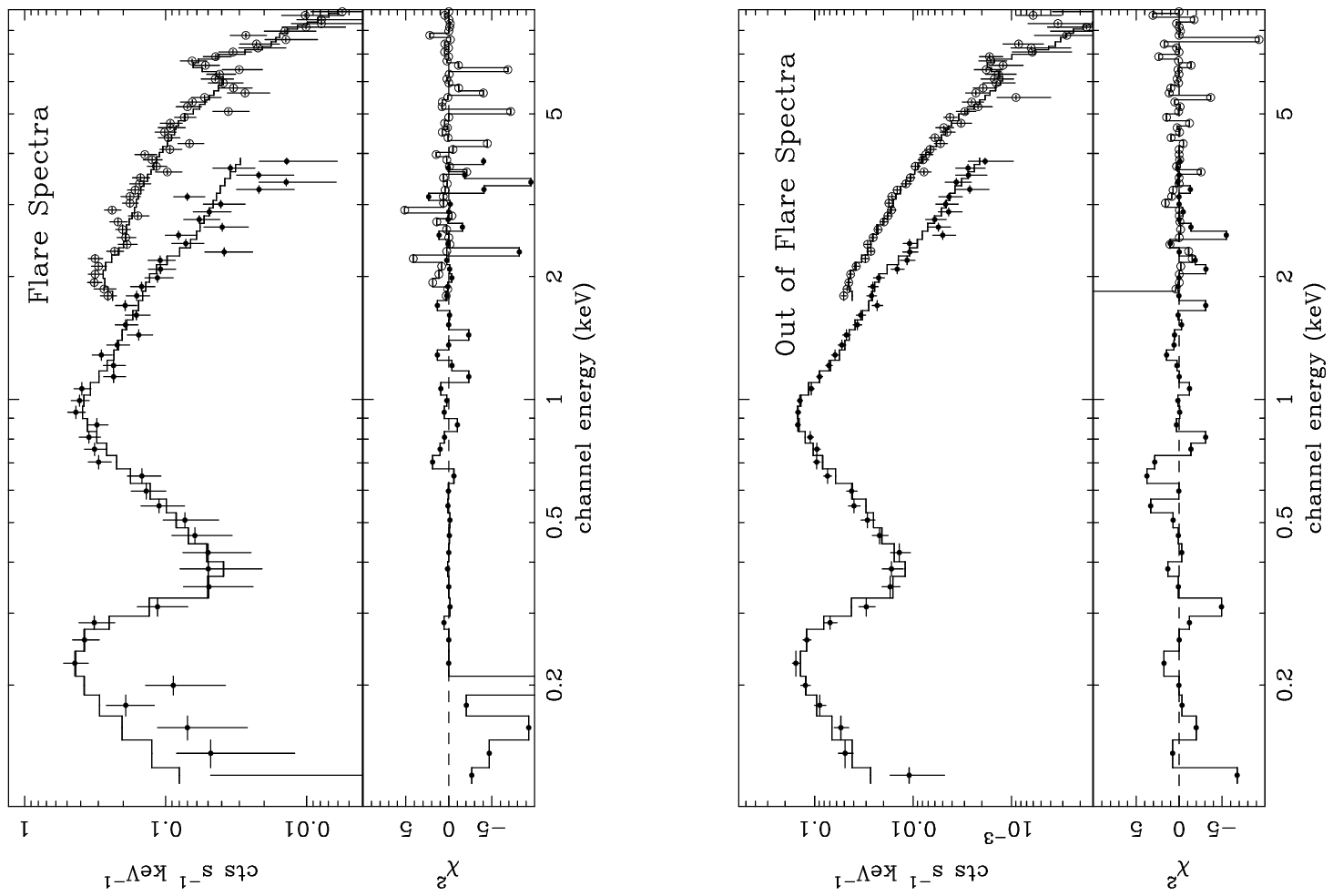}}}}\\
\end{tabular}
\end{center}
\caption{SAX LECS + MECS spectra of \object{HD\,9770}. Spectra were collected 
during the flare (top panel), and outside the flare (bottom panel). In both cases,
the signed contributions to $\chi ^2$ are plotted in the lower panel.}
\label{fig:spec}
\end{figure}

In Fig. \ref{fig:spec} we report the LECS + MECS spectra, and their
best fits for the flare only and for the quiescent 
emission outside the main flare. In both cases it was impossible to fit a 
single temperature model, as expected, or a 2-temperatures model with 
abundances fixed to solar values. Good fits to the data could only be obtained 
with a 2-T model with the metal abundance $Z$ free to vary.
The models shown in Fig.~\ref{fig:spec} refer to the 2-T MK
model; the best-fit parameters together with errors (90\% confidence)
are reported in Table~\ref{tab:spec}.
for both the 2-T MK and RS models. There is no significant difference
between the results obtained with the two plasma codes. 

It can be seen that \object{HD\,9770} is characterized by a hot 
plasma, with values that are normally found for this class of 
sources (e.g. Dempsey et al. \cite{DLFS97}). The Fe K complex at 
$\sim 6.7$\,keV is clearly detected in the
MECS spectrum during the flare, while outside the flare is still present,
although much weaker.
As expected, during the flare the X-ray emission is dominated by hotter 
plasma with a temperature of more than 4 keV. 
Fluxes estimated with the RS and MK model are
reported in Table~\ref{tab:fluxes} for three different bands 
and for both considered datasets. Note that these values are very
similar to the ones detected with ROSAT (Cutispoto et al. 1997).

\section {Discussion}

We have observed the visual triple system \object{HD\,9770} for 42 hours, 
i.e. for 3.7 orbital cycles of the eclipsing binary component present in 
the system. 
Our {\it Beppo}SAX observation confirms that \object{HD\,9770}
is a strong and variable 
coronal source, with most of the emission likely coming from the eclipsing 
binary itself. We have detected significant variability in the quiescent 
emission of the star, plus localized variability in a couple of short-lived 
flares. The variability observed outside the flares can be explained in terms 
of orbital modulation and/or eclipses. In fact these variations are smooth and
detected over more than 3 orbital cycles, thus they should be due primarily
to geometric effects, rather than intrinsic variability. The optical eclipse
at phase zero seems detected also in the X-rays, immediately after a maximum 
in the light curve. In correspondence of 
the secondary eclipse we have no data in the folded light curve due to the 
windowing. Otherwise, if we use all data but the strongest flare, we would 
have a maximum in the flux at phase $\sim 0.5$ due to the small flares seen
during the second orbital cycle and at the end of the observation. 
Thus, although there is not much similarity between the optical and the x-ray
folded light curve besides the eclipse at phase 0, the modulation seems to apply
to the full x-ray light curve. This could indicate that, beside the eclipsing 
effect, there is also a self-eclipsing modulation of the coronal emission. 
Note that Cutispoto et al. (1997) could not find, in the ROSAT all sky survey 
data, any modulation since the sampling was very poor (see Sect.\,1).

The fact that the hardness ratio seems also modulated by the system rotation
is an indication that hotter and cooler plasmas are likely
confined in coronal structures of 
different sizes, with the hotter plasma confined in smaller structures. 
This is the opposite of what has been reported for the RS CVn systems 
\object{AR\,Lac} and \object{TY\,Pyx} on the basis of EXOSAT data 
(White et al. 1990, Culhane et al. 1990). However, 
Ottmann et al. (1993) did not find differences between the rotational
modulation of soft and hard X-rays in a ROSAT observation of \object{AR\,Lac},
rising some doubts about the EXOSAT results. These differences 
could also be explained by coronal variability, but the higher S/N
and larger bandwidth ASCA observation
of White et al. (1994) seems to support the ROSAT results. 
In any case, our data are of low S/N 
and do not allow us to firmly asses the existence of two different 
families of loops with widely different sizes.
Clearly this source should be observed with larger effective area detectors. 

A 2-T thermal model is a good representation of the spectra with a reduced
$\chi^2_r \sim 1$. The low-temperature component has a value of about 0.7 keV
and the high-temperature component a value of about 2 keV. The two components 
have comparable emission measures. During the flare, the high-temperature 
component increases both in temperature and emission measure, while the 
low-temperature component remains essentially unaffected. The average 
temperature throughout the flare is in excess of 4 keV and the emission measure
is a factor $\sim$ 6 higher than during quiescence. These values are consistent
with those typically found in X-ray observations of active stars (Schmitt et 
al. 1990, Dempsey et al. 1993, 1997; Ortolani et al. 1997a, 1997b).

As discussed by Cutispoto et al. (1997), \object{HD\,9770} confirms the 
correlation between L$_{\rm bol}$ and L$_{X}$ found by Pallavicini et al. 
(\cite{PTS90})
for the full sample of UV Ceti and BY Dra-type of stars observed with EXOSAT.
For \object{HD\,9770} this is confirmed both by ROSAT and {\it Beppo}SAX data.
For a discussion of the implications of this point see Cutispoto et al. (1997).

Although we do not, yet, know accurately the photospheric metallicity
of \object{HD\,9770}, from a preliminary spectral synthesis analysis of an
high resolution optical spectrum obtained recently by us at ESO,
we find that [Fe/H]$\sim -0.5$. We are currently doing a more
detailed analysis of these data, however the indication is that the coronal 
abundance of HD\,9770 is consistent with the photospheric value
and that both are subsolar. 
Similar results have also been obtained recently for a {\it Beppo}SAX
observation of the star \object{VY Ari} (Favata et al.\  \cite{FMP97b}).
\object{VY Ari} is an active non
eclipsing SB1 binary with the visible star classified as K3--4/V--VI
(Bopp et al.\  \cite{BSA89}),
i.e. a source very similar to \object{HD\,9770}. 
Subsolar coronal metal abundances are found in many active stars,
both with or without subsolar photospheric metal abundances. This has become 
evident with the study of many ASCA spectra (e.g. 
White et al.\  \cite{WAD94}, White \cite{W96}; Singh et al.\  \cite{SDW95}, 
\cite{SWD96}; Tagliaferri et al.\  \cite{TCF97}, Ortolani et al. 1997a, 
Mewe et al.\  \cite{MKW96b}, \cite{MKO97}), and by other satellites 
(Tsuru et al.\  \cite{TMO89}; Stern et al.\  \cite{SUT92}, \cite{SLS95};
Ottmann \& Schmitt \cite{OS96}; Schmitt et al.\  \cite{SSD96}; Mewe et al.\  
\cite{MKW96b}, \cite{MKO97}).
Thus, it is by now clear that the coronal metal abundances found for most 
of the very active stars are sub-solar. However, it is by no means obvious
that these low coronal metallicities are also in contradiction with the 
measured photospheric abundances in the same stars.
Indeed the low metal abundances found by ROSAT and ASCA for the coronae of
\object{CF Tuc} (Schmitt et al.\  \cite{SSD96}) and \object{$\lambda$\,And}
(Ortolani et al. 1997a), as well as the first results from
{\it Beppo}SAX for \object{Capella} (Favata et al.\  \cite{FMB97a})
and $\beta$ Ceti (Maggio et al. 1998),
do not seem to be in contradiction with their photospheric values.
On the contrary, the low metal abundance found for the young stars
\object{AB Dor} (Mewe et al.\  \cite{MKW96b}, Pallavicini et al.
\cite{PTM99}) and \object{HD\,35850} 
(Tagliaferri et al.\  \cite{TCF97}) are in contradiction with their 
photospheric abundances which are solar. 

Finally we note that the coronal metallicity found from fitting the X-ray
spectra seems to be somewhat higher during the flare than outside of it.
However, the change is not statistically significant, and can only be
considered as a hint for a behavior similar
to that seen previously in strong flares detected in \object{Algol} with
ROSAT (Ottmann \& Schmitt \cite{OS96}) and in \object{AB\,Dor} and
\object{II\,Peg} with ASCA (Mewe et al 1997, Ortolani et al. 1999).
If confirmed, these time dependent abundances would give some support to
the possibility of real abundance differences between the photosphere and 
the corona of some active stars, possibly induced by magnetic activity. 

\acknowledgements{This research was financially supported by the Italian Space 
Agency. We thank the {\it Beppo}SAX Science Data Center (SDC) for their support
in the data analysis. Finally we thank Sofia Randich for communicating 
us her results of a preliminary spectral synthesis analysis of the optical 
spectrum of \object{HD\,9770}.
We also thank the anonymous referee for her/his comments which improved
an earlier version of this paper.}

\end{document}